\documentclass[12pt]{article}
\usepackage[T1]{fontenc}
\usepackage[utf8]{inputenc}
\usepackage[main=english]{babel}

\usepackage{amsmath,amsfonts,amssymb,mathrsfs,eucal,bm}
\usepackage{graphicx}
\usepackage{dsfont}
\usepackage{fancyhdr}
\usepackage{geometry}

\usepackage[final]{microtype}
\usepackage{hyperref}
\hypersetup{
	colorlinks=true,
	linkcolor=blue,
	filecolor=blue,      
	urlcolor=blue
}
\usepackage[nodayofweek]{datetime}

\usepackage{comment}

\usepackage{subcaption}

\providecommand{\U}[1]{\protect\rule{.1in}{.1in}}

\newcommand{\Z}{\ensuremath{\mathbb{Z}}}
\newcommand{\bk}{\ensuremath{\mathbf{k}}}
\newcommand{\bK}{\ensuremath{\mathbf{K}}}
\newcommand{\bW}{\ensuremath{\mathbf{W}}}
\newcommand{\be}{\ensuremath{\mathbf{e}}}
\newcommand{\bx}{\ensuremath{\mathbf{x}}}
\newcommand{\bX}{\ensuremath{\mathbf{X}}}
\newcommand{\bO}{\ensuremath{\mathbf{0}}}
\newcommand{\bv}{\ensuremath{\mathbf{v}}}
\newcommand{\N}{\ensuremath{\mathbb{N}}}
\newcommand{\R}{\ensuremath{\mathbb{R}}}

\newcommand{\bP}{\ensuremath{\mathbb{P}}}
\newcommand{\bS}{\ensuremath{\mathbb{S}}}

\providecommand{\U}[1]{\protect\rule{.1in}{.1in}}

\newtheorem{Conj}{Conjecture}

\begin{document}

\title{Anomalous diffusion properties of stochastic transport by heavy-tailed jump processes}
\author{Paolo Cifani, Franco Flandoli, Lorenzo Marino}
\maketitle

\begin{abstract}
In this work, we investigate the large-scale transport properties of a passive scalar advected by a turbulent fluid, modelled as a superposition of divergence-free vector fields, each weighted by an independent symmetric $\alpha$-stable-like process. Motivated by recent works \cite{Cifani:Flandoli25,Luo:Teng25} showing that complex small-scale spatial structures often lead to Brownian dispersion, we study if this principle persists when the driving noise exhibits heavy-tailed jump statistics. Our numerical results show a clear dichotomy linked with the tail behaviour of the noise. When considering standard $\alpha$-stable processes, very large jumps survive the interaction with the spatial complexity and yield anomalous, super-diffusive transport. In contrast, when the $\alpha$-stable noise is either truncated or exponentially tempered, suppressing extremely long jumps, the transport undergoes a transition to a classical diffusive regime.
\end{abstract}

\section{Introduction}
The problem of the diffusive properties of turbulent transport%
\begin{equation}
\partial_{t}T+\mathbf{u}\cdot\nabla T=0
\label{eq:transport}
\end{equation}
(possibly with a molecular diffusion term $\kappa\Delta T$) is a basic one in
Physics (from the intuition of Boussinesq \cite{Boussinesq77}, used so often in fluid
and plasma models), Engineering and Mathematical Physics \cite{Berselli:Iliescu:Layton06,Majda:Kramer99}. Stochastic models of turbulent velocity fields, of the
form%
\[
\mathbf{u}\left(  \mathbf{x},t\right)  =\sum_{\mathbf{k}}\mathbf{\sigma
}_{\mathbf{k}}\left(  \mathbf{x}\right)  \xi_{t}^{k}%
\]
where $\mathbf{\sigma}_{\mathbf{k}}\left(  \mathbf{x}\right)  $ are vector
fields, $\xi_{t}^{\mathbf{k}}$ are scalar (possibly generalized) stochastic
processes and the sum is extended to suitable wave numbers $\mathbf{k}$, have
been used for long time \cite{Kraichnan94}. In recent years, these models have sparked renewed interest as a suitable description of small-scales of turbulent fluids \cite{Majda:imofeyev:Vanden01,Galeati20,Flandoli:Galeati:Luo22,book:Flandoli:Luongo23,Coghi:Maurelli23,Holm:Luesink:Pan21,Resseguier:Memin:Chapron17,Luo21,Butori:Mayorcas:Morlacchi25,Ephrati:Cifani:Luesink:Geurts23,Ephrati:Jansson:Papini26}, and many more authors and works. Most of the classical works cited above and others consider the case when $\xi_{t}^{k}$ is a white noise (derivative of a Brownian motion) or an
Ornstein-Uhlenbeck process, approximating white noise in the small
time correlation limit. The diffusive properties of such turbulent transport models
have been thoroughly investigated in many of the above mentioned works.

Very recently, there has been an increasing interest in the case when $\xi_{t}^{k}$
are either processes with memory or processes with large jumps. The reason is
the kind of fluid regimes one aims to describe. In 3D turbulence, or quasi-two
dimensional turbulence when the small eddies behave like in 3D, the turbulent
vortex structures are small with short relaxation time, so that an approximation by white noise or Ornstein-Uhlenbeck process with small correlation time looks reasonable. But in the case of 2D turbulence, longer correlations appear and larger vortex structures start to dominate the energetics, due to the inverse
cascade. Therefore the description by white noise or close to white noise
processes may miss important transport information. Up to now, it is not exactly clear which kind of processes is more suitable for describing such phenomena and thus it became important to investigate the diffusion properties of other processes, with memory or long jumps (mimicking long coherent excursions).

Concerning the case where $\xi_{t}^{k}$ are related to Fractional Gaussian
Noises (FGN) with Hurst parameter $H>1/2$ (the persistence case), let us
mention two works: in \cite{Flandoli:Russo23} a theoretical result is proved, which
gives a strong indication of the fact that FGN with $H>1/2$ produces stronger
diffusion than classical white noise. Unfortunately, such result has been obtained in the case $\mathbf{u}\left(  \mathbf{x},t\right)$ has a very simple space structure:  a single field $\mathbf{\sigma}_{\mathbf{k}
}\left(  \mathbf{x}\right)  $ or when the $\mathbf{\sigma
}_{\mathbf{k}}\left(  \mathbf{x}\right)  $ are constant. Therefore it was not clear if the result was stable by space-complexity of $\mathbf{u}\left(  \mathbf{x},t\right)  $, as it happens in turbulent fluids. This was investigated numerically in \cite{Cifani:Flandoli25}, where several small-scale $\mathbf{\sigma}_{\mathbf{k}}\left(  \mathbf{x}\right)  $ fields (similar to those used in the white noise case) multiplied
by independent FGN processes with $H>1/2$ have been studied. The result has
been quite unexpected: the behaviour is similar to the white noise case. A
tracer in such a velocity field behaves like a Brownian motion, since it jumps
from one component to the other of the family of $\mathbf{\sigma}_{\mathbf{k}%
}$ and thus its displacements are governed by independent processes $\xi
_{t}^{k}$.

The present paper aims to investigate the same regime as in \cite{Cifani:Flandoli25} but in the case when $\xi_{t}^{k}$ are $\alpha$-stable processes with large jumps. Namely, we want to see whether the long-tails structure is preserved in the presence of a complex space structure. We shall investigate this problem numerically, after a proper theoretical formulation. The same problem has been studied theoretically recently \cite{Luo:Teng25}, providing a rigorous result on the diffusion properties; but the $\alpha$-stable processes $\xi_{t}^{k}$ in \cite{Luo:Teng25} satisfy a certain condition of small or moderate jumps, different from the one we have in mind; the final result of \cite{Luo:Teng25}, under that condition, is a Brownian behaviour (namely a phenomenon similar to \cite{Cifani:Flandoli25}, but rigorously proved).

We shall investigate three different cases, also motivated by the comparison with the theoretical result of \cite{Luo:Teng25}:
\begin{enumerate}
\item[$(J_1)$] a cylindrical alpha stable noise, namely independent alpha-stable processes along each mode of the space-dependent noise;
\item[$(J_2)$] truncation of the previous alpha-stable processes;
\item[$(J_3)$] exponential tempering of them.
\end{enumerate}
Our results about these three cases are summarized in Section \ref{Subsec:Main}, with the corresponding numerical simulations presented in Section \ref{Sect_numerical}. The processes and equations investigated here are introduced in Sections \ref{Subsec_background} and \ref{Subsec:sol_concept}.

\section{Model formulation and Main results}
\label{Sect model}

Our aim is to investigate numerically a modification of the model appeared in \cite{Cifani:Flandoli25}, in which the original fractional Brownian random weights $B^{H,k}_t$ are replaced by some heavy-tailed (non-Gaussian) noises. More precisely, we would like to understand under which assumptions on the time scales and the considered noise, one should expect an anomalous diffusion in the large wave-number limit, as opposed to the diffusive behaviour observed in \cite{Cifani:Flandoli25} and \cite{Luo:Teng25}.
 
 We are interested in the following transport equation on $\R^2$:
\begin{equation}
	\label{eq:main1}
	\begin{cases}
		\partial_tT+\mathbf{u}\cdot\nabla_xT=0;\\
		T_{|t=0}=T_0,
	\end{cases}    
\end{equation}
where  $\mathbf{u}=(u_1(\bx,t),u_2(\bx,t))$ represents a "turbulent" incompressible fluid. Here, we encode the turbulence as:
\[\mathbf{u}(\bx,t) := uC(\eta,\tau,\alpha)\sum_{\bk\in \bK_\eta}\sigma_\bk(\bx)\dot{Z}^{\alpha,\bk}_t,\]
where $u$ is the \emph{average velocity} constant (with dimension $[L]/[T]$), $C(\eta,\tau,\alpha)$ is a normalizing constant whose precise expression will be given later, $\eta$ indicates the \emph{space scale} with dimension $[L]$ and the parameter $\tau$ appearing in $C(\eta,\tau,\alpha)$ represents the \emph{relaxation time} of the fluid with dimension $[T]$. The index set $\bK_\eta$ amounts for the wave numbers $\bk\in \Z^2$ which are activated at a fixed space scale $\eta>0$ and it is given by
\[\bK_\eta:= \left\{\bk\in \Z^2\colon |\bk|\in \left[\frac{1}{2\eta},\frac{1}{\eta}\right] \right\}.\]
Note in particular that the limit for $\eta\to 0$ corresponds indeed to the large wave number limit. The divergence-free  vector field $\sigma_k$ are dimensionless and defined as
\[\sigma_\bk(\bx):=\begin{cases}
	\sqrt{2}\cos(\bk\cdot \bx)\frac{\bk^\perp}{|\bk|}, \mbox{ if }\bk\in \bK^+_\eta\\
	\sqrt{2}\sin(\bk\cdot \bx)\frac{\bk^\perp}{|\bk|}, \mbox{ if }\bk\in \bK^-_\eta
\end{cases}\]
where $\bK^+_\eta$ is the set of $\bk = (k_1, k_2) \in \bK_\eta$ such that either ${k_1 > 0}$ or $\{k_1 = 0, k_2> 0\}$, and $\bK^-_\eta= -\bK^+_\eta$.
 
Above, each $\dot{Z}^{\alpha,\bk}_t$ denotes the formal time derivative of $(Z^{\alpha,\bk}_t)_{t\ge 0}$, a symmetric "$\alpha$-stable-like" Lévy process on $\R$. Such processes are characterized by discontinuous trajectories. The parameter $\alpha\in (1,2)$, known as the \emph{stability index}, quantifies the frequency and relevance of large jumps: smaller values of $\alpha$ correspond to dynamics dominated by rare but very large jumps whereas in the formal limit as $\alpha\to 2$, one recovers the continuous paths of a Brownian motion. More precisely, we will test our model under three possible scenarios:
\begin{enumerate}
\item[$(J_1)$] $(Z^{\alpha,\bk}_t)_{t\ge 0}$ are independent, standard $\alpha$-stable processes;
\item[$(J_2)$] $(Z^{\alpha,\bk}_t)_{t\ge 0}$ are independent, truncated $\alpha$-stable processes with \emph{cut-off} $\epsilon>0$; 
\item[$(J_3)$] $(Z^{\alpha,\bk}_t)_{t\ge 0}$ are independent, exponentially tempered $\alpha$-stable processes with \emph{tempering parameter} $A>0$. 
\end{enumerate}
The above cases represent "prototypical" examples of jump-driven noises and well illustrate how variations in the intensity and/or frequency of the jumps influence the model limit behaviour. Roughly speaking, a standard $\alpha$-stable process exhibits a power-law jump statistics: jumps of arbitrarily large size may occur with a frequency determined by a power law distribution.  Its dimension is $[T]^{1/\alpha}$, which is suggested by its $\alpha$-self-similarity. In case $(J_2)$, one starts again from the standard $\alpha$-stable process above but then removes all jumps larger than the cut-off $\epsilon>0$. Finally, the exponentially tempered $\alpha$-stable process can be viewed as an interpolation between the two previous models: it retains jump activity at all scales but large jumps become exponentially rare, with a rate controlled by the parameter $A$. In both cases $(J_2)$ and $(J_3)$, the dynamics preserves heavy-tail features at small scales, while exhibiting Gaussian-like behaviour at large ones. This suggests that the dimension of $\dot{Z}^{\alpha,\bk}_t$ is given by $[T]^{(\alpha-1)/\alpha}$. Hence, a dimensional analysis implies that the normalizing constant $C(\eta,\tau,\alpha)$ must have dimension $[T]^{(\alpha-1)/\alpha}$. More precisely, it is given by
\[C(\eta,\tau,\alpha)=\left[\textit{Card}(\bK_\eta)\right]^{-1/\alpha}\tau^{(\alpha-1)/\alpha}.\]
Note that when $\eta$ is small enough, $\textit{Card}(\bK_\eta)\sim \pi\frac{1}{\eta^2}-\pi\frac{1}{4\eta^2}=\pi\frac{3}{4\eta^2}$.

We are thus interested in the following SPDE on $\R^2$:
\begin{equation}
	\label{eq:main}
	\begin{cases}
		dT+ uC(\eta,\tau,\alpha)\sum_{\bk\in \bK_\eta}\sigma_\bk(\bx)\cdot\nabla_xT\diamond d{Z}^{\alpha,\bk}_t=0;\\
		T_{|t=0}=T_0,
	\end{cases} 
\end{equation}
The symbol $\diamond$ above means the Marcus stochastic integral which is a generalization of the more well-known Stratonovich integral for discontinuous paths processes. As the Stratonovich one, the Marcus integral incorporates in its definition some corrector terms in order to recover the standard differentiation properties. 

Precise definitions of the above stochastic processes, together with the meaning of all parameters, and the corresponding stochastic integration theory, are provided below.

\subsection{Some backgrounds on stable Lévy processes and their modifications}
\label{Subsec_background}

Let $(\Omega, \mathcal{F},\bP)$ be a probability space and $d\in \N$ a generic dimension.
An $\R^d$-valued stochastic process
 $(Z_t)_{t\ge 0}$ is called a \emph{Lévy process} if it starts from zero almost surely, has independent and stationary increments, and is
càdlàg (i.e.\ it has right continuous paths having left limits $\mathbb{P}$-almost surely).  A fundamental tool in the analysis of Lévy processes is given by the Lévy-Kintchine formula (see for instance \cite{book:Jacob05}). It allows to represent
the Lévy symbol $\Phi$ of $(Z_t)_{t\ge0}$, given by
\[\mathbb{E}[e^{i\xi \cdot Z_t}] \, = \, e^{t\Phi(\xi)}, \quad \xi \in \R^d\]
in terms of the generating triplet $(b,\Sigma,\nu)$ as:
\[\Phi(\xi) \, = \, i b \cdot \xi -\frac{1}{2} \Sigma\xi\cdot \xi+\int_{\R^d_0}\bigl(e^{i \xi\cdot z}-1-i \xi\cdot z \mathds{1}_{B(0,1)}(z)\bigr)
\,\nu(dz), \quad \xi \in \R^d,\]
where $b$ is a vector in $\R^d$, $\Sigma$ is a symmetric, non-negative definite matrix in $\R^d\otimes \R^d$ and $\nu$ is a Lévy measure on
$\R^d_0:=\R^d\smallsetminus \{0\}$, i.e.\ a $\sigma$-finite measure on $\mathcal{B}(\R^d_0)$, the Borel $\sigma$-algebra on $\R^d_0$, such
that $\int(1\wedge \vert z\vert^2) \, \nu(dz)$ is finite.
In particular, the law of any Lévy process is completely determined by its generating triplet $(b,\Sigma,\nu)$. In this work, we restrict our attention to \emph{symmetric pure-jump Lévy} processes, namely processes for which $\Sigma=0$, $b=0$ and the Lévy measure $\nu$ is symmetric.

The first class (i.e.\ case $(J_1)$) of noises considered in this work consists of $\alpha$-stable Lévy processes, with stability index $\alpha\in(1,2)$. These are symmetric pure-jump Lévy processes whose increments follow an $\alpha$-stable distributions. Their Lévy measure takes the form: 
\[\nu_\alpha(dz) :=\mu(d\theta)\frac{dr}{r^{d+\alpha}},\qquad z=r\theta \in \R^d, \, \theta \in \bS^d, \, r>0\] 
where $\mu$ is a finite Borel measure  on $\bS^d$, the unitary sphere on $\R^d$, known as the \emph{spectral measure} of $Z_t$. In particular, we will talk about
an  \emph{isotropic} $\alpha$-stable process when its spectral measure is rotational invariant. In such a case, its characteristic function reduces to
\[\mathbb{E}[\exp(i Z_t\cdot\xi)] = \exp\left[-\sigma_\alpha |\xi|^\alpha t\right] , \qquad \xi \in \R,\]
 for some constant $\sigma_\alpha>0$, called  the \emph{scale parameter}.
Indeed, if $Z_t$ is an isotropic $\alpha$-stable process with scale parameter $\sigma_\alpha$, then for any $\lambda>0$, the process $\lambda Z_t$ remains isotropic $\alpha$-stable, with scale parameter $\sigma_\alpha\lambda^\alpha$.
In the particular case where $\sigma_\alpha=1$, the corresponding process is usually called a \emph{standard (isotropic)} $\alpha$-stable process.

As already noted above, $\alpha$-stable processes are self-similar with index $1/\alpha$, i.e.\ $(Z_{ct})_{t\ge 0}\overset{d}{=} (c^{1/\alpha}Z_t)_{t\ge 0}$ for any $c>0$. Unlike the Gaussian case, $\alpha$-stable processes do not possess finite moments of all orders. More precisely, they admit finite fractional moments only up to order $\beta<\alpha$. In the isotropic case with scale parameter $\sigma_\alpha$, one can actually obtain an explicit formula (cf.\ \cite[Eq.\ (25.6)]{book:Sato99}) for $\beta\in (0,\alpha)$:
\begin{equation}
	\label{eq:1}
\mathbb{E}[|Z_t|^\beta] = \frac{2^\beta\Gamma(\frac{d+\beta}{2})\Gamma(1-\frac{\beta}{\alpha})}{\Gamma(\frac{d}{2})\Gamma(1-\frac{\beta}{2})}(\sigma_\alpha t)^{\beta/\alpha}.
\end{equation}

A fundamental distinction from the Brownian case is that understanding the global behaviour of an $\alpha$-stable process cannot be achieved by examining only its projections onto the coordinate axes. This limitation arises from the fact that, in general, the components of the process are not independent. Consequently, $\alpha$-stable processes possess a substantially richer and more intricate geometric structure. In order to analyse the geometry of the support of the associated spectral measure, it is therefore necessary to consider projections along arbitrary directions (cf.\ \cite[Thm.\ 2.1.5, Ex.\ 2.3.4]{book:Samorodnitsky:Taqqu94} and \cite[Prop.\ 11.10]{book:Sato99}). If $(Z_t)_{t\ge 0}$ is an $\alpha$-stable process with spectral measure $\mu$ and $v\in \bS^d$ is a direction, then $Z_t\cdot v$ is again an $\alpha$-stable process (on $\R$) with scale parameter
\[\sigma_\alpha(v)=\int_{\bS^d}|v\cdot \theta|^\alpha \, \mu(d\theta).\]
On the other hand, if for any $v\in \bS^d$, $Z_t\cdot v$ is an $\alpha$-stable process with scale parameter $\sigma_\alpha(v)$ and such a constant is actually independent from $v$, then one can conclude that  $Z_t$ is an isotropic $\alpha$-stable process on $\R^d$ with scale parameter $\sigma_\alpha$.

Starting from a given Lévy process, one can construct new Lévy processes by means of several standard modifications, such as linear transformations or subordination. In this work, we focus on two classical examples of such modifications, namely truncation and exponential tempering, which suppress large jumps while preserving $\alpha$-stable behaviour at small scales.

A truncated $\alpha$-stable process $(Z_t)_{t\ge 0}$ is obtained by removing large jumps from its Lévy measure. Given a cut-off parameter $\epsilon>0$, it has the form:
\[\nu(dz):= \mathds{1}_{\{|z|\le \epsilon\}}\nu_\alpha(dz),\qquad z \in \R^d,\]
where $\nu_\alpha$ is an $\alpha$-stable Lévy measure.

An exponentially tempered $\alpha$-stable process modifies the Lévy measure by introducing an exponential damping of large jumps. For a tempering parameter  $A>0$, the Lévy measure is
\[\nu(dz):=e^{-A|z|}\nu_\alpha(dz),\qquad  z\in \R^d,\]
where $\nu_\alpha$ is again some $\alpha$-stable Lévy measure.

One of the key property of tempered and truncated $\alpha$-stable processes is that, unlike $\alpha$-stable process, they have finite moments. 

\subsection{Solution concept for transport equation}
\label{Subsec:sol_concept}

In connection with the method of characteristics, it is natural in this context to interpret the stochastic transport equation \eqref{eq:main} in the sense of Marcus stochastic integration (cf.\ \cite{Hartmann:Pavlyukevich23}). Indeed, Itô interpretation is inappropriate in this context, as it does not preserve the classical chain rule and introduces artificial drift corrections unrelated to the transport mechanism. On the other hand, while Stratonovich integration provides a consistent framework for stochastic transport equation driven by continuous path noises (cf.\ \cite{Cifani:Flandoli25}), it does not correctly describe the dynamics induced by discontinuous trajectories. Indeed, in the presence of jumps, Stratonovich integral models the effect of each jump as a purely instantaneous displacement, without taking into account the infinitesimal flow generated by the vector field during the jump. By contrast, the Marcus integral accounts for the internal dynamics of jumps by associating each jump of the driving process with the deterministic flow generated by the vector field. Such interpretation preserves the geometric structure of the transport equation and it is consistent with the characteristic method. Informally, a solution $T$ to \eqref{eq:main} is characterized as
\[T(\bX^\eta_t(\bx),t) = T_0(\bx),\]
where $\bX^\eta_t(\bx)$ is the flow map, solution to the characteristics equation:
\begin{equation}
\label{eq:char}
\begin{cases}
	d\bX^\eta_t(\bx)=uC(\eta,\tau,\alpha)\sum_{\bk\in \bK_\eta}\sigma_\bk(\bX^\eta_t(\bx)) \diamond d{Z}^{\alpha,\bk}_t;\\
	\bX^\eta_0(\bx)=\bx.
\end{cases}
\end{equation}

The symbol $\diamond$ above means the Marcus stochastic integral which is a generalization of the more well-known Stratonovich integral for discontinuous paths processes. For a precise formulation of this integral, see \cite[Section 6.10]{book:Applebaum09}. Under our assumptions, i.e.\ smooth, divergence-free vector fields $\sigma_\bk$ and symmetric pure-jump Lévy drivers with finite first moment (i.e.\ $\alpha>1$ or truncated/tempered variants), it is well-known that the characteristic Equation \eqref{eq:char} admits a unique global càdlàg solution defining a stochastic flow of $C^1$-diffeomorphisms almost surely (see, again \cite[Section 6.10]{book:Applebaum09}). For the precise connection between the stochastic transport Equation \eqref{eq:main} and the associated characteristics Equation \eqref{eq:char} under our assumptions, as well as the well-posedness of such models, see \cite{Hartmann:Pavlyukevich23}.

\subsection{Main Results}
\label{Subsec:Main}

We here outline a brief description of the main results. They will be described in more details in Section \ref{Sect_numerical}. Roughly speaking, our numerical results show that the large-scale transport behaviour is strongly determined by how rare extremely long jumps are in the driving noise. When the velocity field is driven by a standard $\alpha$-stable process (i.e.\ case $(J_1)$), large jumps survive the interaction with the complex spatial structure and lead to anomalous, super-diffusive transport: particle displacements scale like $t^{1/\alpha}$, and the limiting dynamics is consistent with an $\alpha$-stable process. In contrast, when large jumps are either truncated or exponentially tempered (i.e.\ cases $(J_2)$ and $(J_3))$, their effect is progressively suppressed, and the tracer dynamics shift to a classical diffusive regime. In these cases, after a short transient period dominated by jump-like behaviour, the particle motion becomes Gaussian with the standard $t^{1/2}$ mean displacement scaling, similarly to what is observed for white noise-driven models. Our results highlight how the tail behaviour of the driving noise, i.e.\ the presence or absence of a power-law jump distribution, plays a crucial role in selecting between anomalous and classical diffusive regimes.

\begin{Conj}
\label{main_thm}
Let $\bx\in \R^2$. Then, as $\eta$ goes to zero, the processes $(\bX^\eta_t(\bx))_{t\ge 0}$ converge in law to the $2$-dimensional process $(\bx+\overline{\bX}_t)_{t\ge 0}$, where:
\begin{itemize}
\item[($J_1$):] $(\overline{\bX}_t)_{t\ge 0}$ is an isotropic $\alpha$-stable process;
\item[($J_2$):] $(\overline{\bX}_t)_{t\ge 0}$ is a centred Brownian motion;
\item[($J_3$):] $(\overline{\bX}_t)_{t\ge 0}$ is a centred Brownian motion.
\end{itemize}
\end{Conj}

The validity of the above result remains an open question but we hope our formulation will motivate further theoretical and practical investigations.
Only in case $(J_2)$, Luo and Teng in \cite{Luo:Teng25} obtained a theoretical result of the diffusive scaling limit. As a preliminary contribution, we provide the following numerical verification. 

In the previous work \cite{Cifani:Flandoli25}, the main objective was to characterize the rate of information diffused by the velocity field, quantified via the mean squared displacement (MSD) of the particle position. While such a metric remains applicable for cases $(J_2)$ and $(J_3)$, it is unsuitable for case $(J_1)$. Indeed, as the $\alpha$-stable driving noise inherently lacks a finite second moment, the resulting particle variance is expected to diverge. In order to compare the three cases on the same ground, we propose instead to analyse the dynamics of the first absolute moment:
\[t\to \mathbb{E}\left[|\bX_t^\eta(\bO)|\right].\]

As an intermediate step toward the validity of Conjecture \ref{main_thm}, our numerical results show that under $\alpha$-stable driving noise (i.e.\ case $(J_1)$), the particle trajectory exhibits super-diffusive behaviour characterized by the scaling $\mathbb{E} \left[|\bX_t^\eta(\bO)| \right]\sim t^{1/\alpha}$. As explained in \eqref{eq:1}, this is the behaviour usually associated to $\alpha$-stable processes. Conversely, for the truncated and tempered noise regimes, we observe that the process converges to classical diffusive behaviour, namely $\mathbb{E}\left[|\bX_t^\eta(\bO)|\right]\sim t^{1/2}$, after a short transient period.

While such information already points towards Conjecture \ref{main_thm}, we provide additional confirmation by computing the probability density function (PDF) of the particle displacement along the $x$-axis at time $t = 1$. By visually comparing their graph with the expected ones, we show that
\begin{itemize}
	\item[($J_1$):] the distribution $\mathcal{L}$ is $\alpha$-stable;
	\item[($J_2$):] the distribution $\mathcal{L}$ is centred Gaussian;
	\item[($J_3$):] the distribution $\mathcal{L}$ is centred Gaussian.
\end{itemize}
where $\mathcal{L}$ be the probability distribution function of $\bX^{\eta}_1(\bO)\cdot \be_1$.

The above claim already yields strong evidence in support of the diffusive limits stated in Conjecture \ref{main_thm} for cases $(J_2)$ and $(J_3)$. 
Consequently, we restrict our subsequent analysis to case $(J_1)$. To better understand the properties of the limiting process, we investigate the geometric structure of the associated spectral measure.

More precisely, we fix various directions $\bv \in \bS^2$ and compare the distribution of the tracer projected along these directions: $\bX^{\eta}_1(\bO)\cdot \bv$. The numerical results presented below indicate that the limiting process is isotropic.

Furthermore, as part of a more detailed analysis, we also aim to estimate the coefficients $\overline{\sigma}$ appearing in front of the limiting processes.

\begin{Conj}\label{main_thm2}
\label{main_rmk}Let $\bx\in \R^2$. Then, as $\eta$ goes to zero, the processes $(\bX^\eta_t(\bx))_{t\ge 0}$ converge in law to the $2$-dimensional process $(\bx+\overline{\sigma}\overline{\bX}_t)_{t\ge 0}$, where:
\begin{itemize}
\item $(\overline{\bX}_t)_{t\ge 0}$ is a standard $\alpha$-stable process in case $(J_1)$ and a centred Brownian motion in the other ones;
\item the scaling parameter is given by
\begin{equation}
\label{eq:sigma_lim}
\overline{\sigma}\sim  \lambda u^{\alpha/H}\eta^{1-\alpha/H}\tau^{(\alpha-1)/H},    
\end{equation}
where, as before, $H=\alpha$ in case $(J_1)$ and $H=2$ in the other cases and the parameter $\lambda>0$ depends on the features of the model (i.e.\ $A$, $\epsilon$, $\sigma_\bk$).
\end{itemize}
\end{Conj}

The heuristic derivation of the scaling parameter $\overline{\sigma}$ is detailed in Appendix \ref{Sect:Appendix} and relies on a suitable time discretization of the tracer dynamics.
As expressed above, the  "approximate" formula for $\overline{\sigma}$ depends however on an additional parameter $\lambda>0$. Below in Section \ref{Sect_numerical}, such a parameter is computed from numerical simulations and then compared against the theoretical prediction. We have not found an argument to predict the coefficient $\lambda$, but we can show numerically that there exists a value providing a good fit between this formula and numerical experiments.

\paragraph{Remark} It appears plausible that, in order to obtain a mathematically rigorous scaling limit, similar to the ones obtained in \cite{Galeati20,Luo:Teng25}, an additional assumption on the interplay between the parameters of the model is required. More precisely, one may assume that the mean velocity $u:=u(\tau,\eta)$ is chosen in such a way that
\begin{equation}
	\label{eq:cond}
	u^{\alpha/H}\eta^{1-\alpha/H}\tau^{(\alpha-1)/H} \sim 1, \qquad \text{as $\eta\to 0$},
\end{equation}
where $H=\alpha$ in case $(J_1)$ and $H=2$ in cases $(J_2)-(J_3)$. 
This scaling condition is designed to ensure that the scaling parameter $\overline{\sigma}$ of the limit process is independent from $\eta$ and thus non-trivial, in the limit as $\eta$ goes to $0$. Without such balance condition, one would not expect the emergence of a meaningful macroscopic transport behaviour since the limiting dynamics would either collapse to a degenerate regime or diverge.

\section{Numerical simulations}
\label{Sect_numerical}

In this section we perform numerical simulations of transport equation (\ref{eq:transport}) with advection velocity $\mathbf{u}(\mathbf{x},t)$ given by the stochastic model \ref{eq:main}. Analogously to our related work \cite{Cifani:Flandoli25}, a Monte Carlo method is employed where particle trajectories are simulated by numerical integration of the equations of characteristics (\ref{eq:char}). In all the results presented below, stochastic differential equations are interpreted in the sense of Marcus integrals and solved numerically via integration of the induced flow ODE for jumps. Expected values and probabilities are approximated by appropriate ensemble averages over $10^4$ realisations.

We consider the following three cases for the stochastic process $Z_t$:
\begin{enumerate}
\item[($J_1$):] $\Delta Z_t \sim \Delta t^{1/\alpha}  S_\alpha(1,0,0)$ sampled from the Chambers-Mallows-Stuck algorithm (See \cite[Sect.\ 1.7]{book:Samorodnitsky:Taqqu94} for additional discussions on the topic);
\item[($J_2$):] $\Delta Z_t$ sampled from the tempered distribution by the Bauemer-Merchaer algorithm (see \cite{Kawai:Masuda11}  for the details on the simulation); 
\item $\Delta Z_t$ sampled from the truncated distribution by a Gaussian approximation of the small jumps (see \cite{Asmussen:Rosinski01} for the details of the algorithm used for simulation). 
\end{enumerate}

\begin{figure}[hbt!]
\centering
\begin{subfigure}[b]{0.45\textwidth}
\includegraphics[width=1\textwidth]{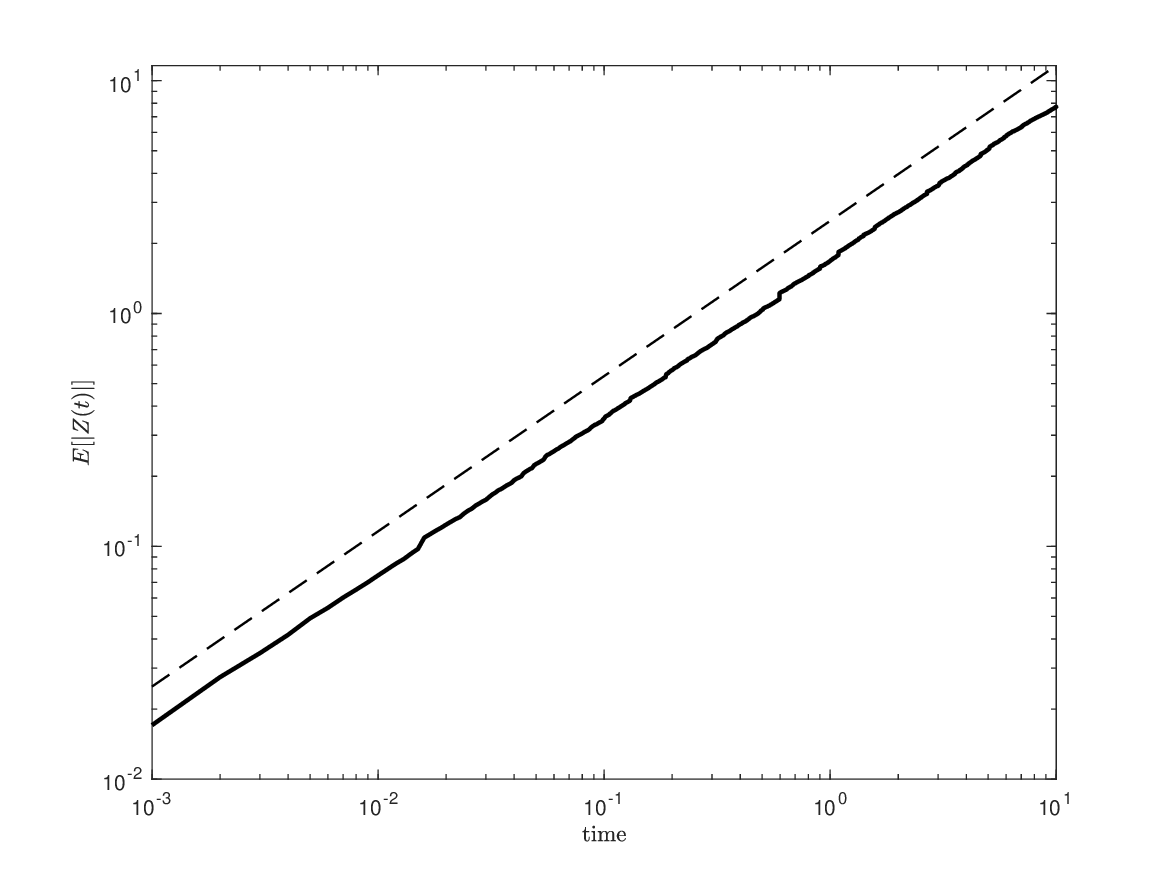}
\end{subfigure}
\hfill
\begin{subfigure}[b]{0.45\textwidth}
\includegraphics[width=\textwidth]{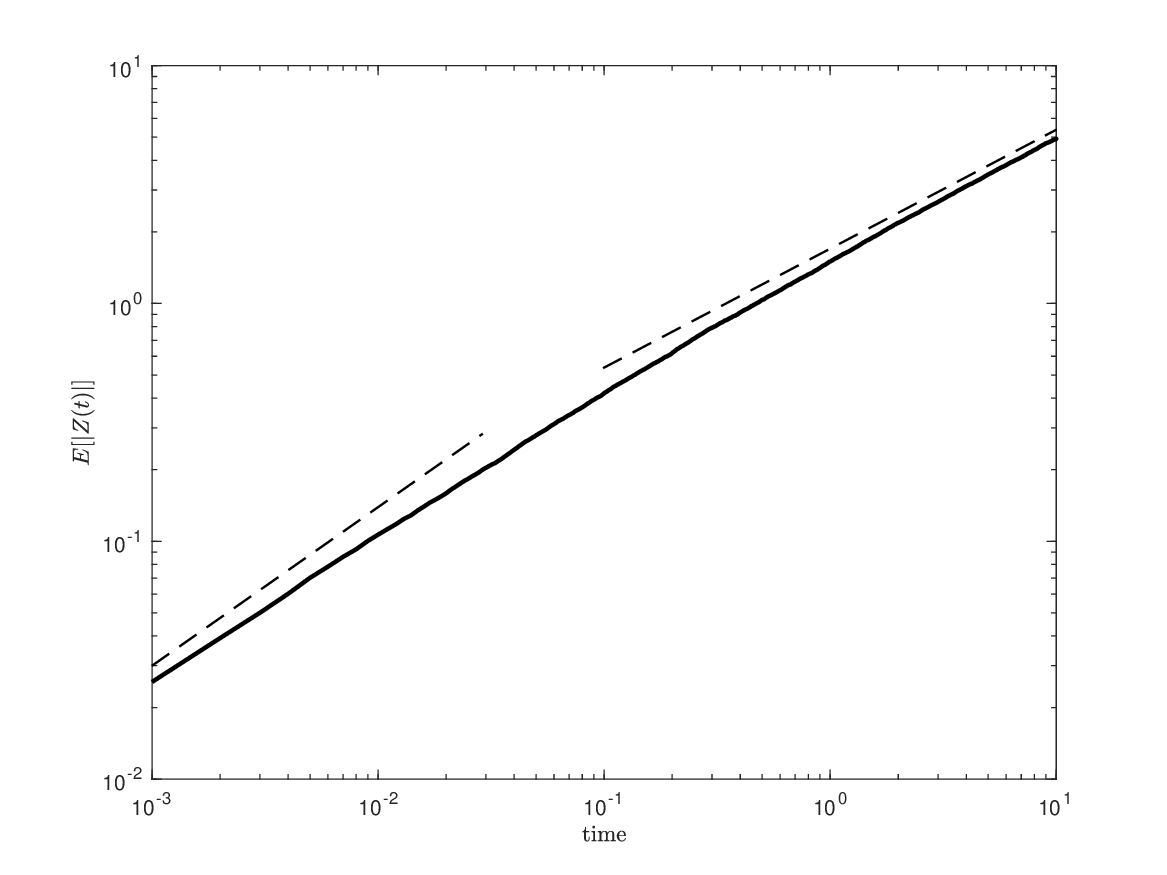}
\end{subfigure}
\hfill
\begin{subfigure}[b]{0.45\textwidth}
\includegraphics[width=\textwidth]{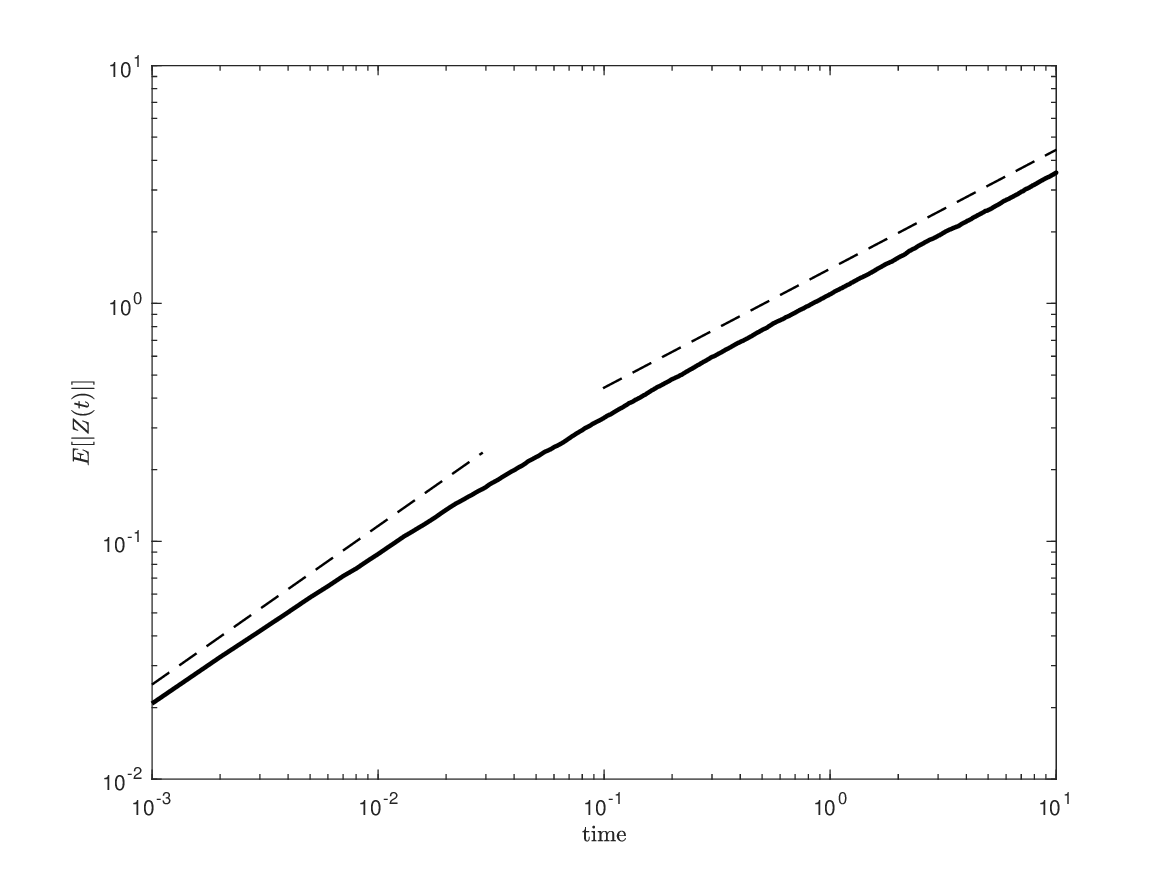}
\end{subfigure}
\caption{$\mathbb{E}[|Z_t|]$ as a function of time for the $\alpha$-stable process (top-left figure), the tempered process (top-right figure) and the truncated process (bottom figure). For the tempered process we set $A=0.3$; for the truncated process we set $\epsilon=10^{-3}$. Slopes $1/\alpha$ and $1/2$ are represented by the dashed lines.}
\label{fig:Z_t}
\end{figure}

As a preliminary step, we show the scaling of $\mathbb{E}[|Z_t|]$ as a function of time for the aforementioned cases. Exponential tempering is adopted to construct an effective Lévy measure whose tails are tamed by a factor $e^{-A}$. Strong jumps are become more and more rare for large values of $A$. In the truncated measure a parameter $\epsilon$ is introduced to suppress all jumps of size larger than $\epsilon$. This behaviour is summarised in Fig. \ref{fig:Z_t}. 

\begin{figure}[hbt!]
\centering
\includegraphics[width=0.7\textwidth]{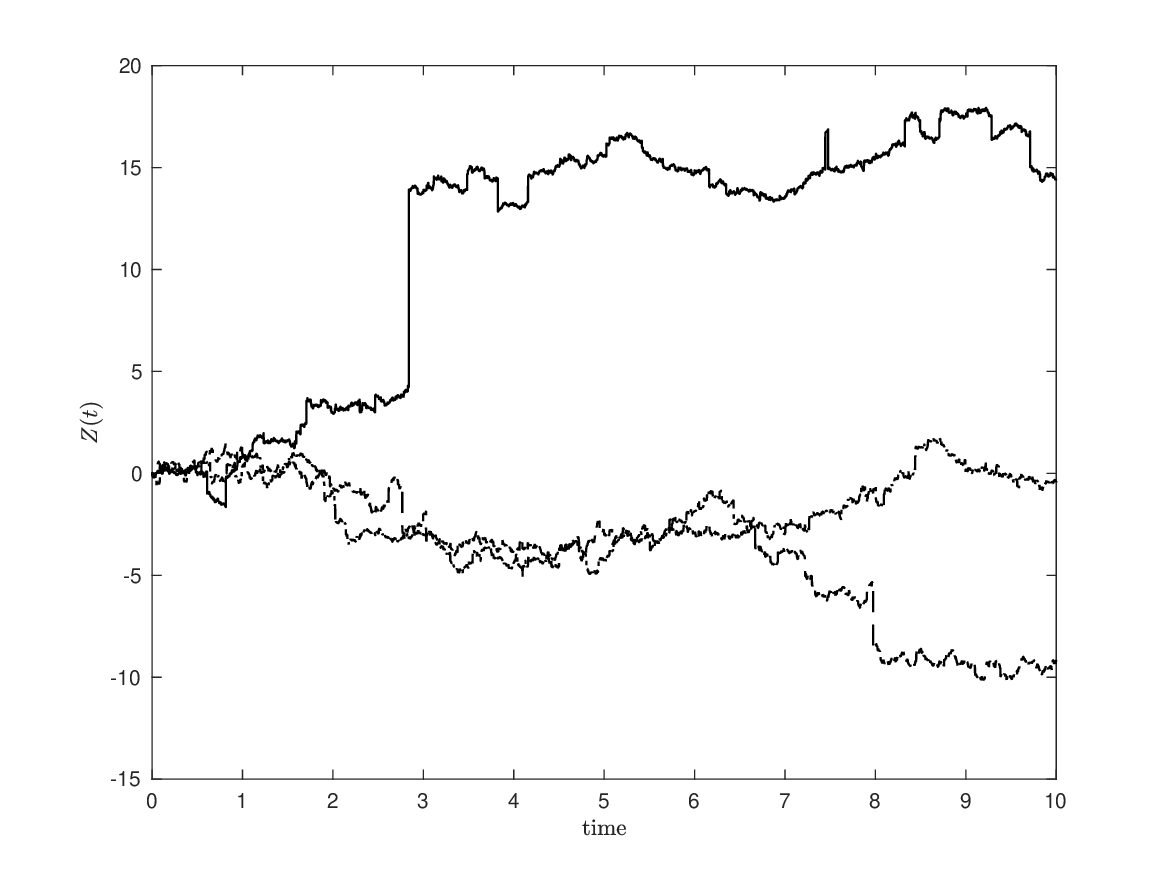}
\caption{Sample paths of the $\alpha$-stable process (solid line), the tempered process (dashed line) and the truncated process (dot-dashed line).}
\label{fig:Zt_path}
\end{figure}

Clearly, the scaling $t^{1/\alpha}$ is shown by the $\alpha$-stable process at all times, as expected from theory. For the tempered and the truncated process, after an initial power law with exponent $1/\alpha$, a transition to Gaussianity takes place with its characteristic scaling $t^{1/2}$ at large enough $t$. The time at which the latter scaling takes over depends on the parameter $A$ and $\epsilon$ of the tempered and truncated measure, respectively. The main observation here is that for small $t$, the process is dominated by jumps while for large $t$ rare events are either tamed or cut out such that diffusion is recovered. An illustration of the sample paths of $Z_t$ is shown in Fig. \ref{fig:Zt_path}.

\begin{figure}[hbt!]
\centering
\begin{subfigure}[b]{0.45\textwidth}
\includegraphics[width=1\textwidth]{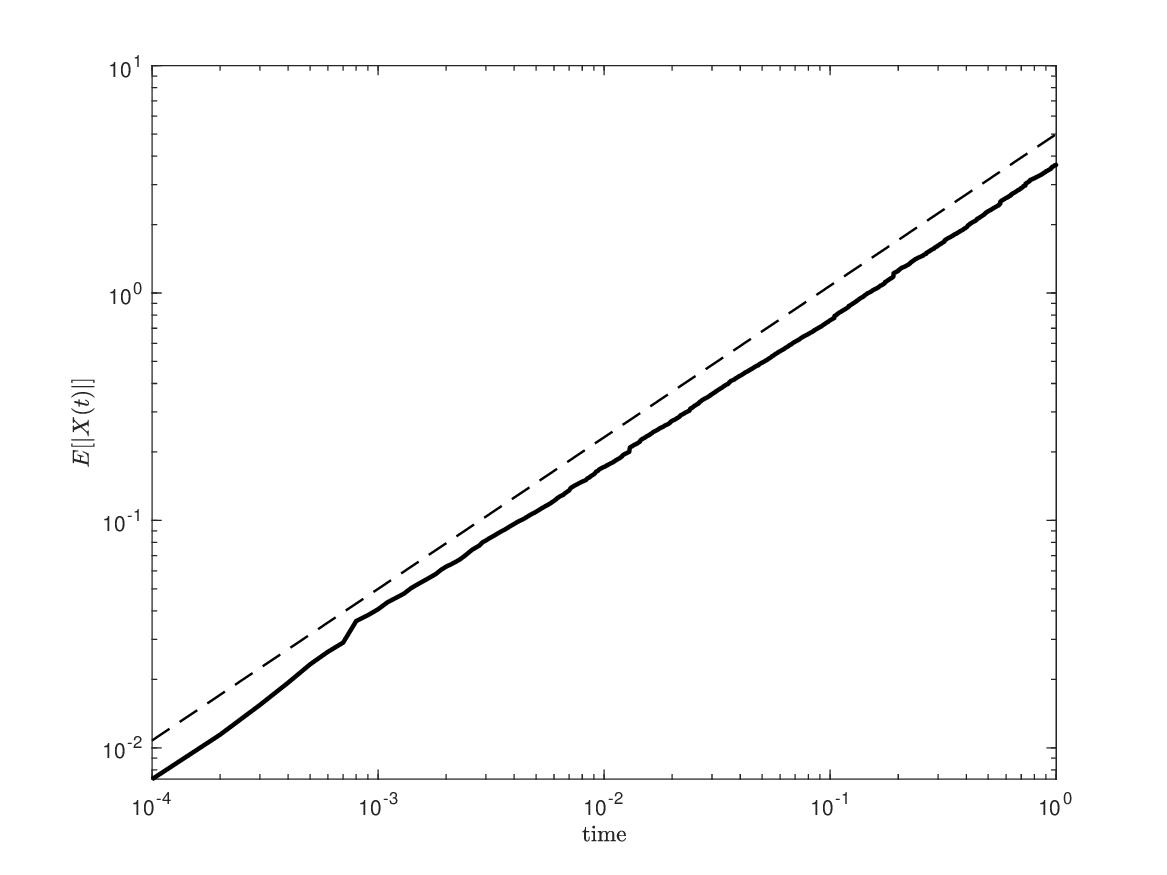}
\end{subfigure}
\hfill
\begin{subfigure}[b]{0.45\textwidth}
\includegraphics[width=\textwidth]{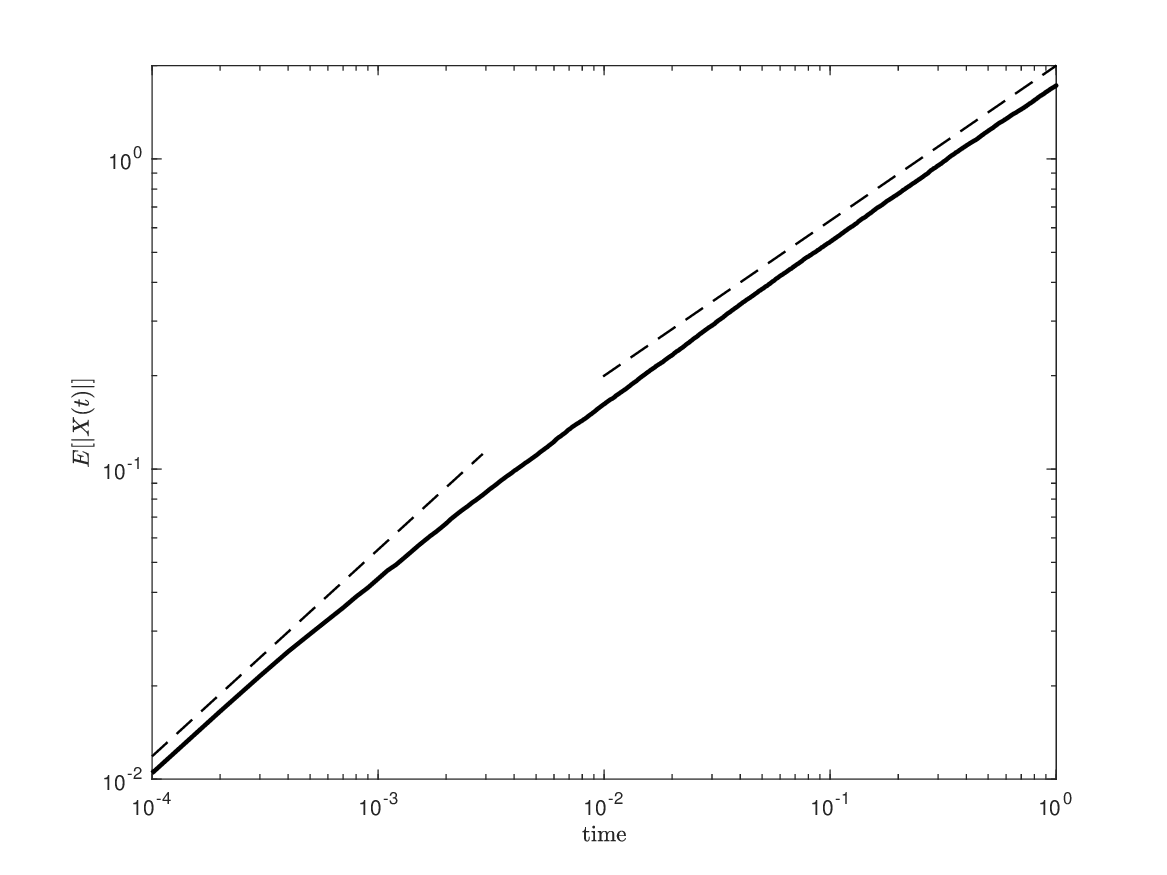}
\end{subfigure}
\hfill
\begin{subfigure}[b]{0.45\textwidth}
\includegraphics[width=\textwidth]{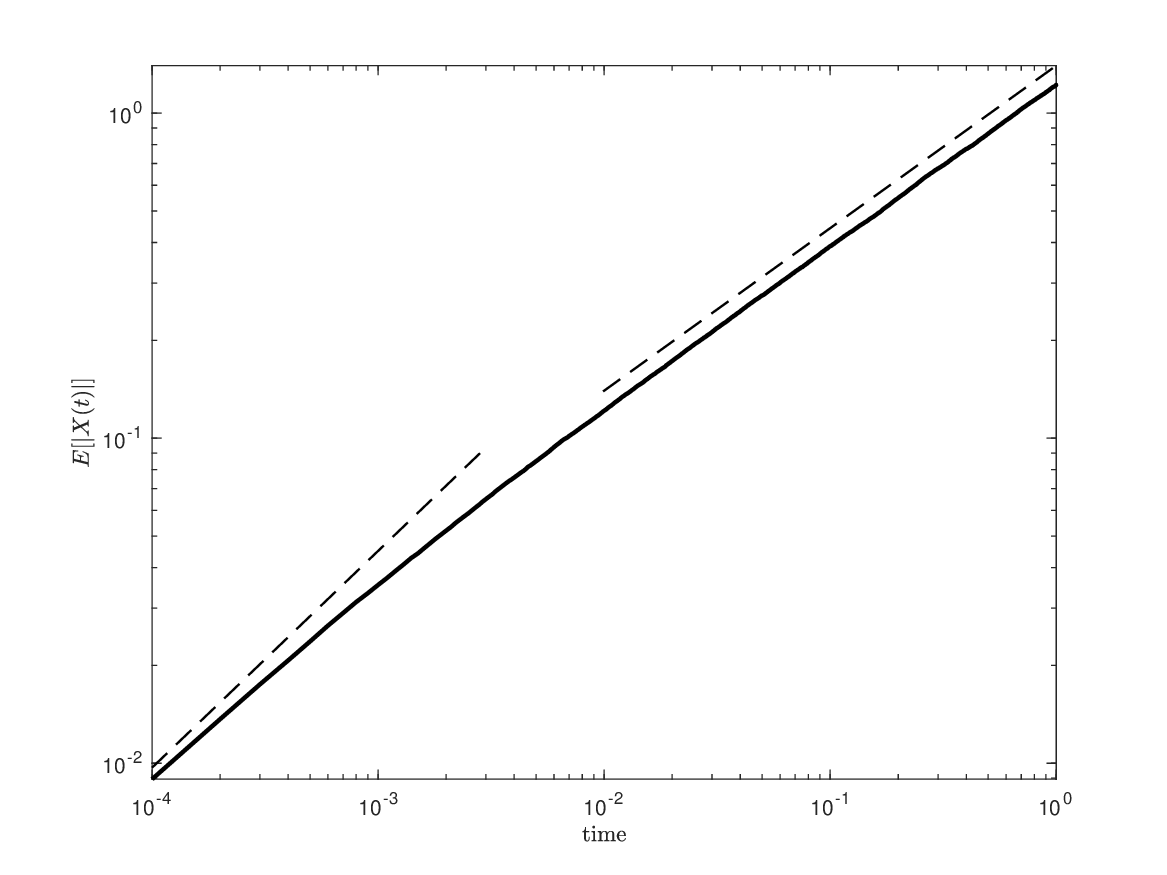}
\end{subfigure}
\caption{$\mathbb{E}[|X_t|]$ as a function of time for the $\alpha$-stable process (top-left figure), the tempered process (top-right figure) and the truncated process (bottom figure). The parameters of the driving stochastic processes are the same as in Fig. \ref{fig:Z_t}. Slopes $1/\alpha$ and $1/2$ are represented by the dashed lines.}
\label{fig:X_t}
\end{figure}

Next, we simulate particle trajectories by numerical integration of (\ref{eq:char}). The reference length scale $\eta$ is set to $2 \pi/20$ and the total simulation time to $T=1$. We consider the same three cases detailed above, namely the driving stochastic processes are the $\alpha$-stable process and its tempered and truncated counterpart. Fig. \ref{fig:X_t} shows $\mathbb{E}[|X_t|]$ as a function of time. 

\begin{figure}[hbt!]
\centering
\includegraphics[width=0.7\textwidth]{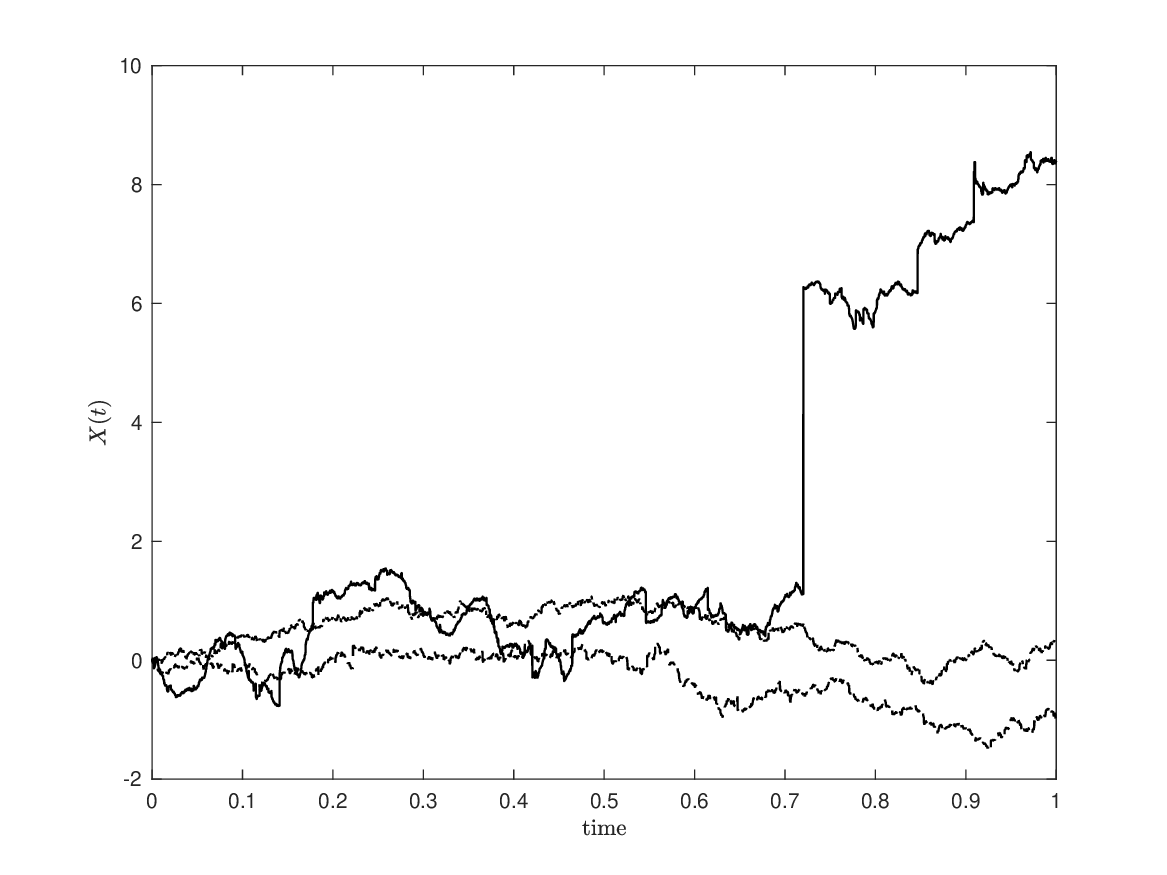}
\caption{Sample paths of $X_t$ for driver $Z_t$ being an $\alpha$-stable process (solid line), a tempered process (dashed line) and a truncated process (dot-dashed line).}
\label{fig:Xt_path}
\end{figure}

Overall the particle position $X_t$ behaves as the driver stochastic process $Z_t$. When the tempered or the truncated processes are used in the vector field, the same transition from the scaling $t^{1/\alpha}$ at small times to the scaling $t^{1/2}$ at large times is observed. We remark here that typical time scale at which this transition takes place is entirely related to the choice of the tempering parameter $A$ and of the truncation threshold $\epsilon$ and not to the physical time $t^*$, defined in \cite{Cifani:Flandoli25}, and linked to the length $\eta$. What is remarkable here is, rather, the fact that $X_t$ maintains the same scaling law of $Z_t$ when the latter is an $\alpha$-stable process, in spite of the complex interactions among the vector field components $\sigma_{\mathbf{k}}$. An illustration of the sample paths of $X_t$ is shown in Fig. \ref{fig:Xt_path}.

\begin{figure}[hbt!]
\centering
\begin{subfigure}[b]{0.45\textwidth}
\includegraphics[width=1\textwidth]{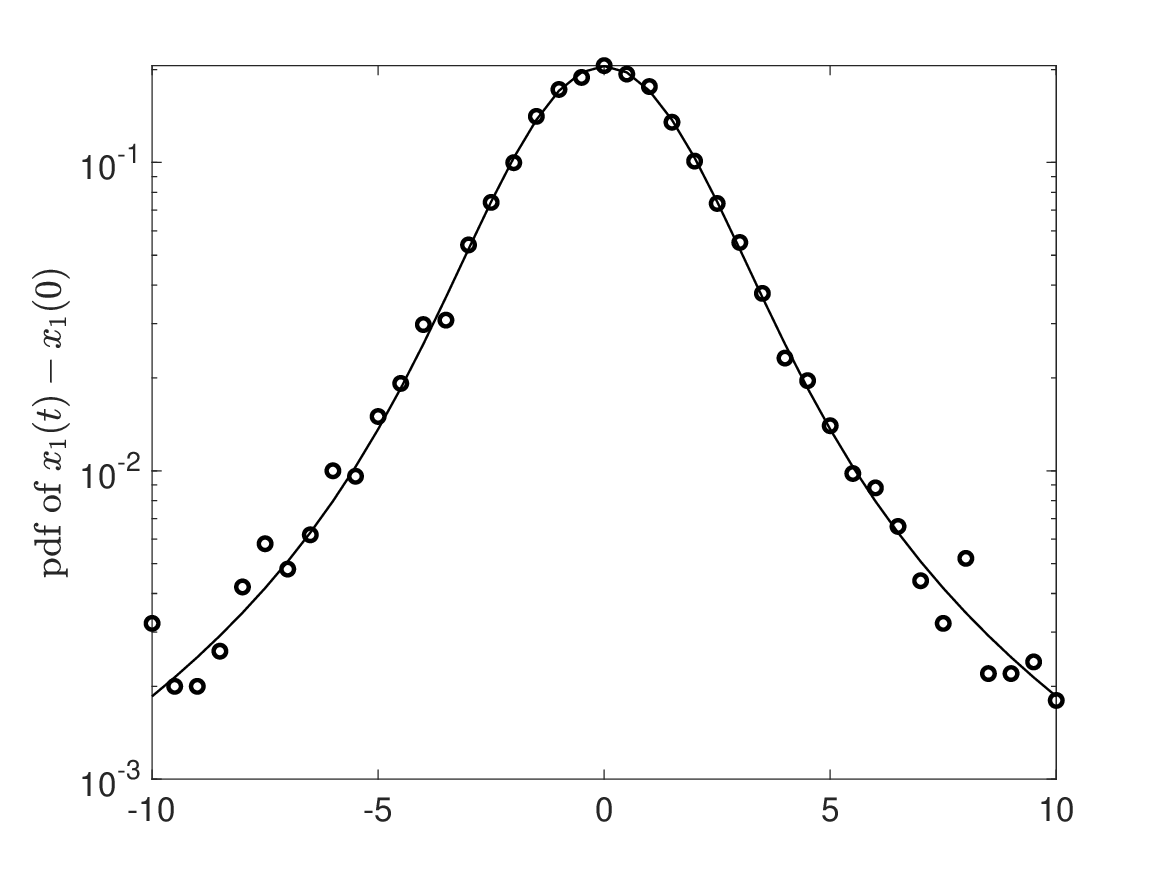}
\end{subfigure}
\hfill
\begin{subfigure}[b]{0.45\textwidth}
\includegraphics[width=\textwidth]{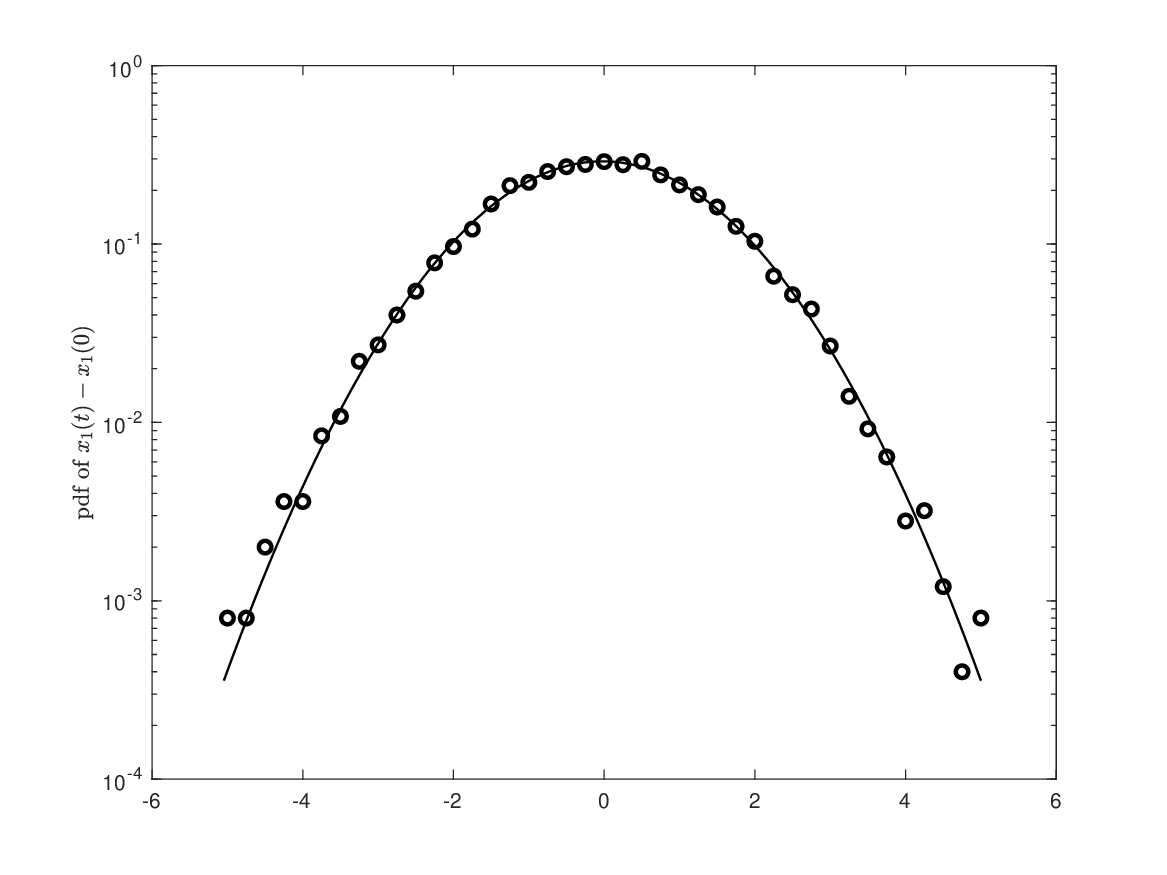}
\end{subfigure}
\hfill
\begin{subfigure}[b]{0.45\textwidth}
\includegraphics[width=\textwidth]{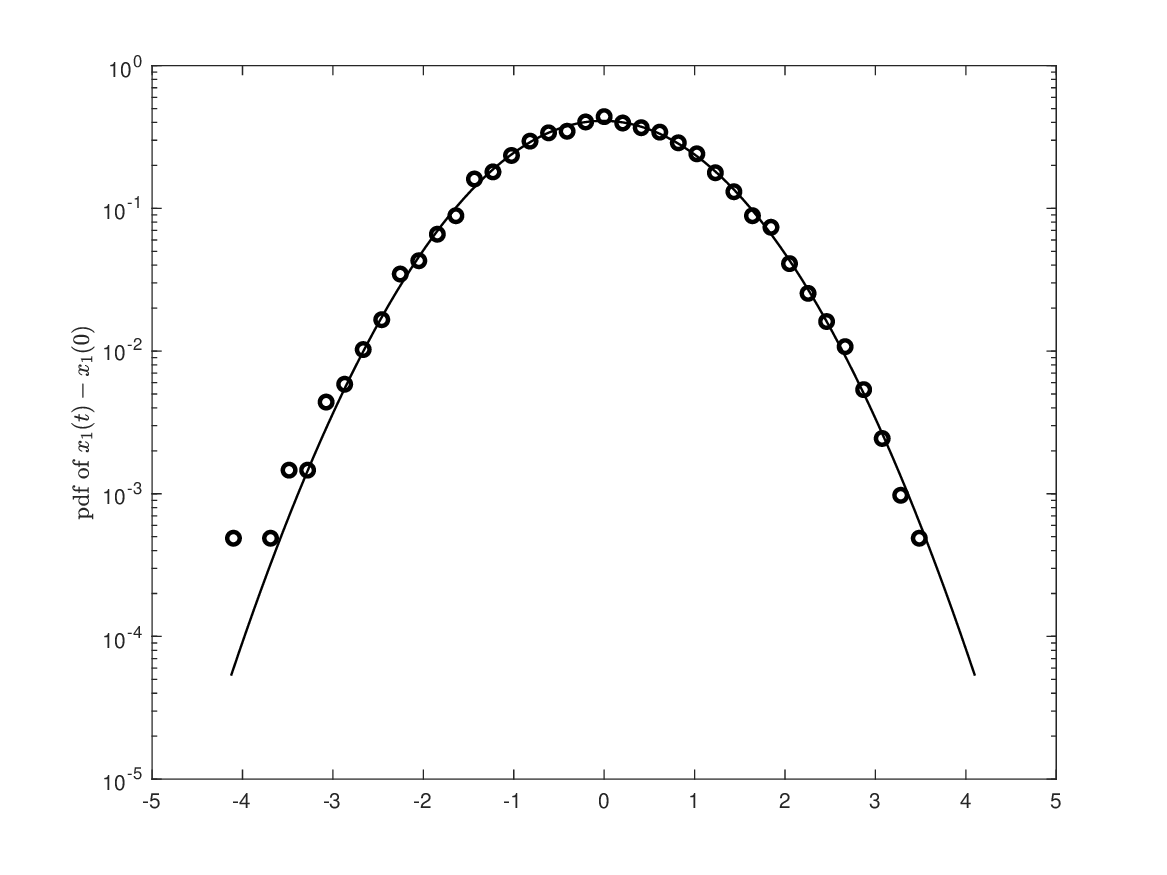}
\end{subfigure}
\caption{Probability distribution function of $X_t$ at the final time $t=1$ for driver $Z_t$ being an $\alpha$-stable process (top-left figure),for driver $Z_t$ being a tempered process (top-right figure) and for driver $Z_t$ being a truncated process (bottom figure).}
\label{fig:pdf_Xt}
\end{figure}

Scaling of the first absolute moment of $t^{1/\alpha}$ and $t^{1/2}$ suggest $\alpha$-stable and Gaussian displacement of the particle $X_t$, respectively. As a further confirmation of this hypothesis, we present in Fig. \ref{fig:pdf_Xt} the probability distribution function of $X_t$ at the final time $t=1$. 
When driving $X_t$ by either the tempered of the truncated process, the computed pdf values (dots) closely adhere to the Gaussian (solid line) having the same mean and standard deviation. Differently, when $Z_t$ is $\alpha$-stable the numerical pdf approximate an $\alpha$-stable distribution with parameter $\sigma$ numerically computed from $\mathbb{E}[|X_t|]$.

To support Conjecture \ref{main_thm}, we compute the pdf of $X_t$, driven by the $\alpha$-stable process, projected along the direction $e_\theta$ for $\theta = \pi/6, \ \pi/3, \ \pi/2$.
\begin{figure}[hbt!]
\centering
\begin{subfigure}[b]{0.45\textwidth}
\includegraphics[width=1\textwidth]{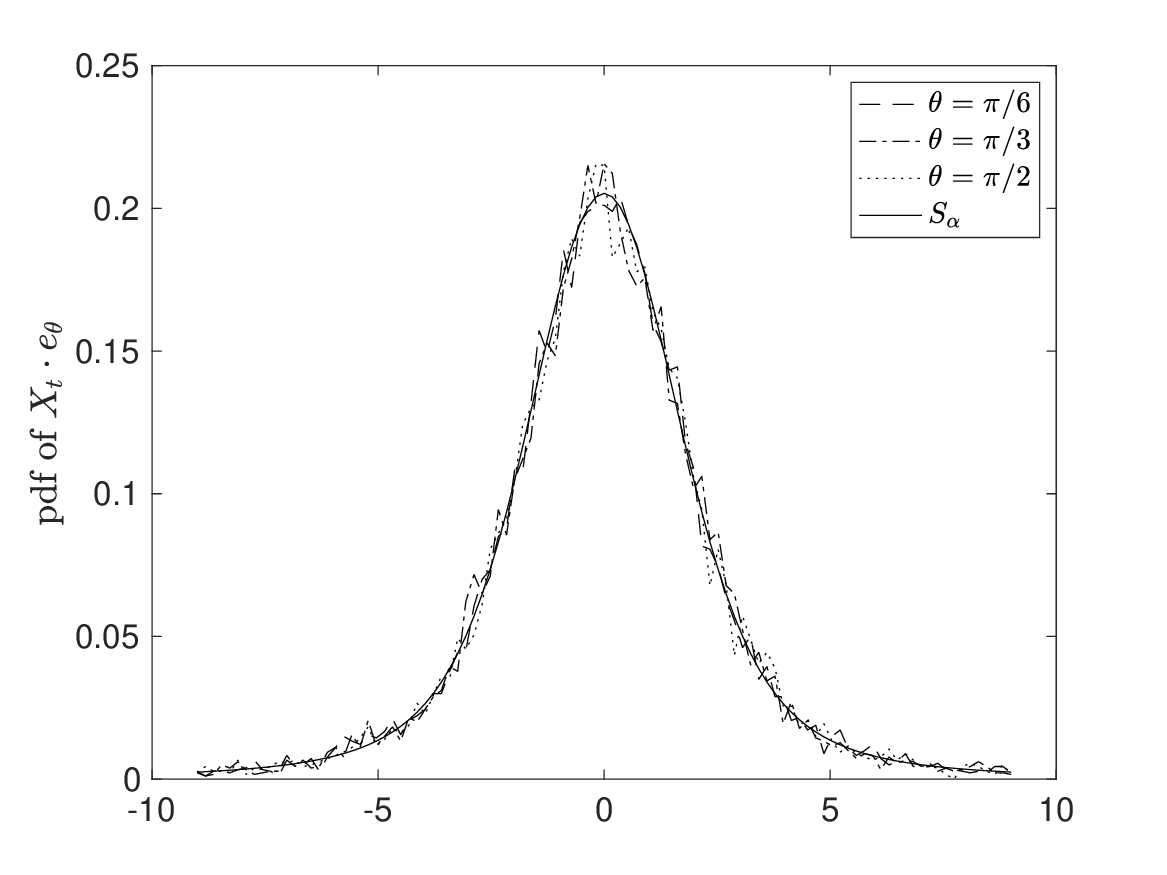}
\end{subfigure}
\hfill
\begin{subfigure}[b]{0.45\textwidth}
\includegraphics[width=\textwidth]{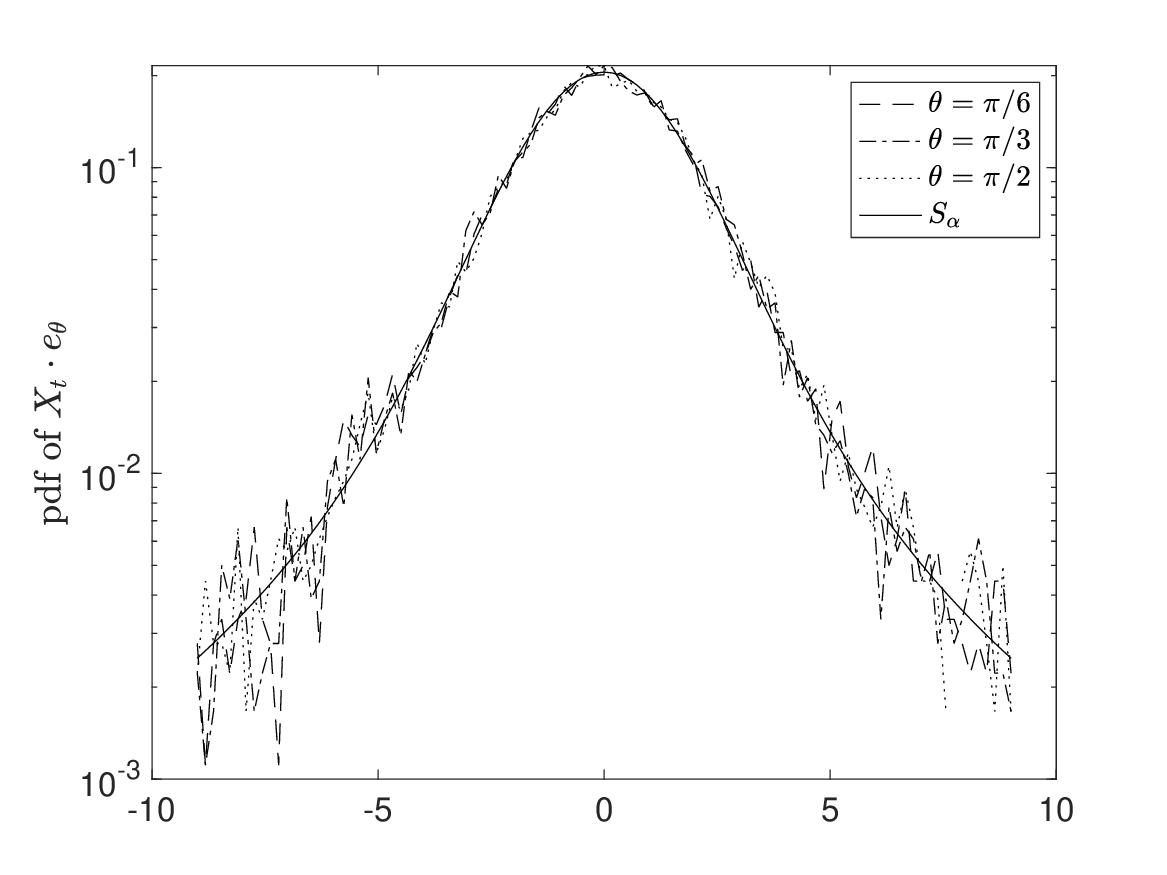}
\end{subfigure}
\caption{Probability distribution function of $X_t \cdot e_\theta$, at the final time $t=1$, using an $\alpha$-stable driver $Z_t$ for $\theta = \pi/6, \ \pi/3, \ \pi/2$. The solid black line is a reference $S_{1.5}$ distribution.}
\label{fig:pdf_Xt_theta}
\end{figure}
As shown in Fig. \ref{fig:pdf_Xt_theta}, at all values of $\theta$ the numerical pdfs closely match the $\alpha$-stable distribution $S_{1.5}$.

\begin{figure}[hbt!]
\centering
\begin{subfigure}[b]{0.45\textwidth}
\includegraphics[width=1\textwidth]{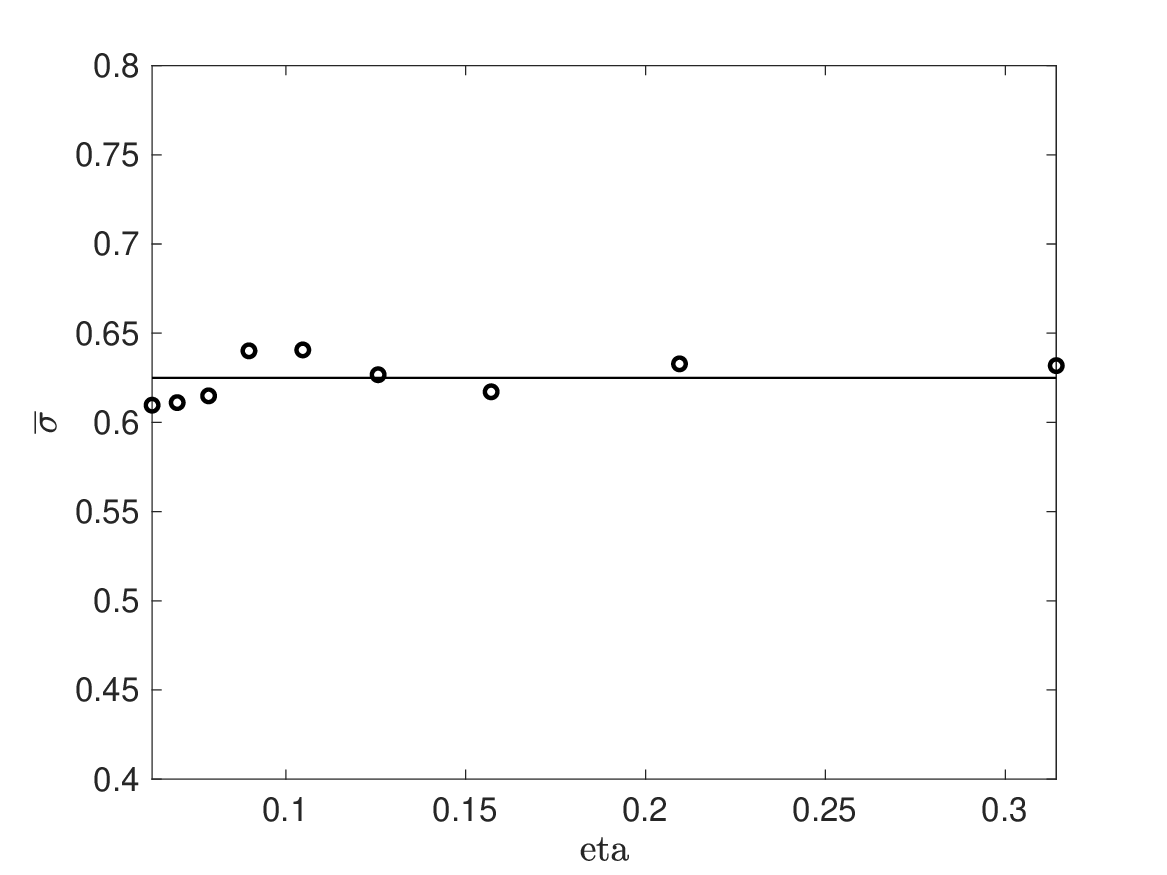}
\end{subfigure}
\hfill
\begin{subfigure}[b]{0.45\textwidth}
\includegraphics[width=\textwidth]{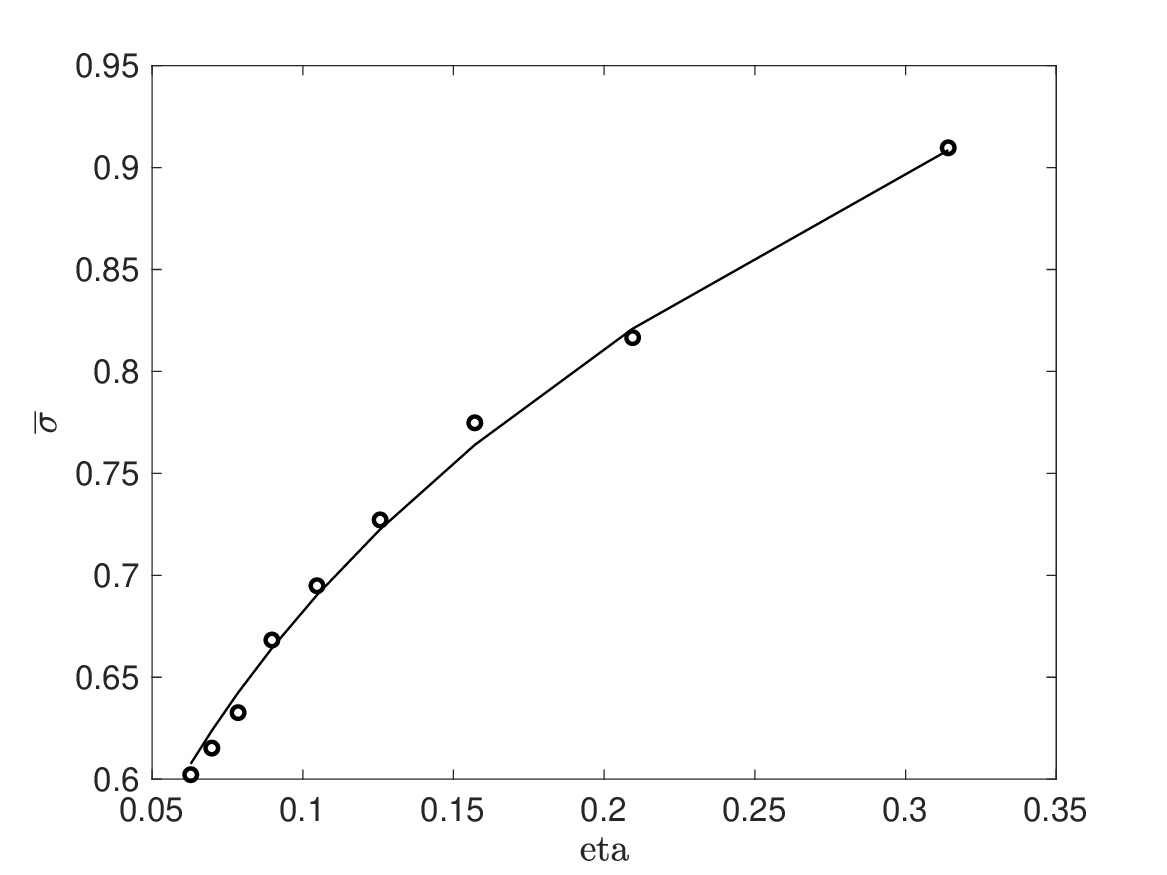}
\end{subfigure}
\hfill
\begin{subfigure}[b]{0.45\textwidth}
\includegraphics[width=\textwidth]{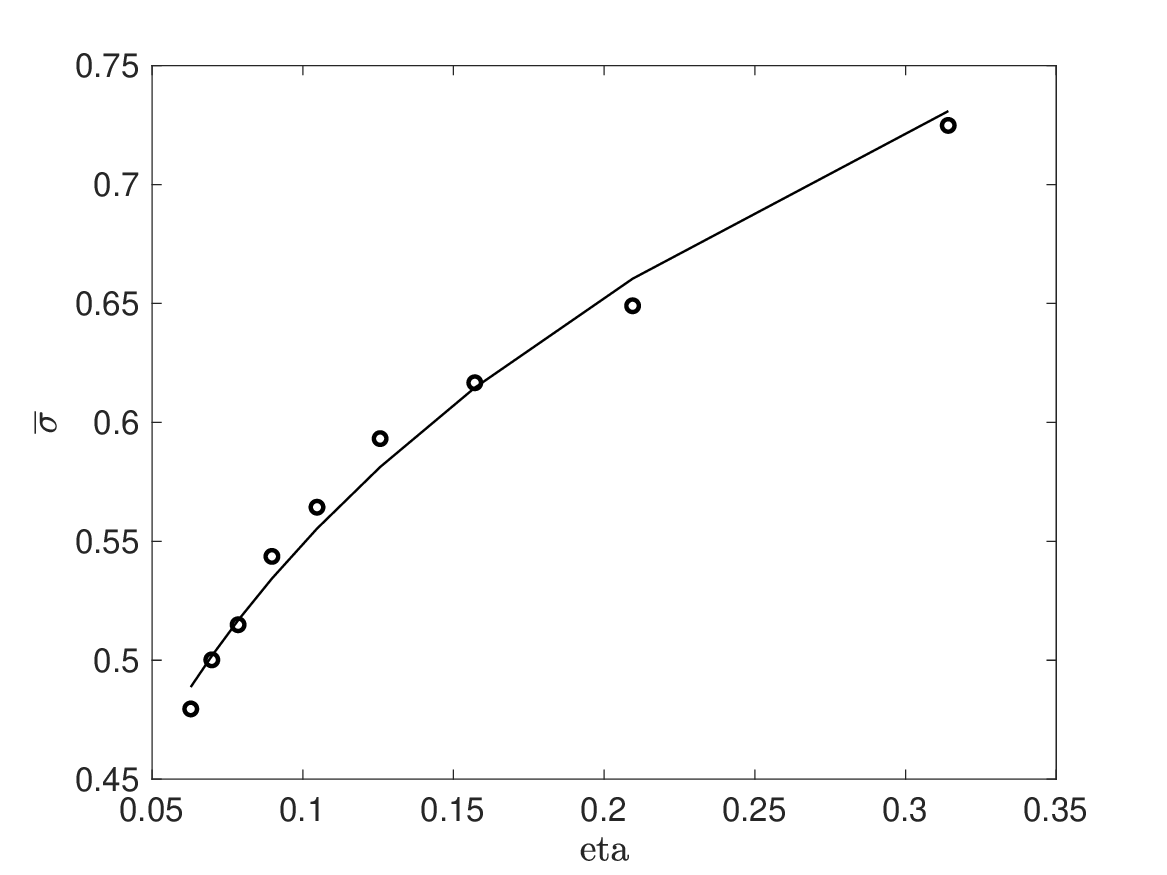}
\end{subfigure}
\caption{Parameter $\overline{\sigma}$ as a function of $\eta$ for driver $Z_t$ being an $\alpha$-stable process (top-left), a tempered one (top-right) or a truncated one (bottom), computed numerically (dots) and analytically (solid line) from \eqref{eq:sigma_lim}. The coefficients of the solid line are fitted from the numerical data.}
\label{fig:sigma_barra}
\end{figure}

As stated in Conjecture \ref{main_thm2}, the truncated and the tempered noise regimes converge to the classical (Brownian) diffusion behaviour at sufficiently large time while the pure $\alpha$-stable case remains stable. An expression for the scaling parameter $\overline{\sigma}$ is provided in (\ref{eq:sigma_lim}). A numerical verification of the validity of the latter would require a series of simulations in the parameter space $u,\eta,\tau$. Furthermore, the parameter $\lambda$ depends as well on parameters of the driving noise, such as the tempering and truncation factors $A$ and $\epsilon$. While an analysis of this kind is in principle possible, it becomes quickly computationally unfeasible. This is the reason why in this work, we limit ourselves to the verification of the dependence of $\overline{\sigma}$ on the length scale $\eta$. Fig.\ \ref{fig:sigma_barra} depicts $\overline{\sigma}$ as a function of $\eta$ in the three noise regimes. A good agreement is found between the numerical values and the postulated theoretical function, further strengthening our claim.

\section{Conclusions\label{Sect Conclusions}}

In this work we have investigated the diffusion effects of certain stochastic models of turbulent fluids, a problem of major importance for applications like the anomalous diffusion in confined fusion plasma.

These stochastic models, also in the works quoted below, are made of the superposition of divergence-free highly oscillating vector fields, each weighted by an independent stochastic process. The difference between the various papers is mostly in the choice of stochastic processes. 

A main technical motivation for this work comes from two previous papers, \cite{Cifani:Flandoli25,Luo:Teng25}. In both of them it is shown that a classical Brownian behaviour of the diffusion arises in spite of the fact that the time-structure of the stochastic turbulence model were not Brownian, but fractional Brownian with $H>1/2$ in one case and $\alpha$-stable (with a bound on the tails) in the second one. In the present work we have shown that this is not always the case.

Our main result is that, if the driving processes of the turbulence model exhibit heavy-tailed jump statistics, then this property persists for the passive scalar advection. The "Brownianization" observed and proved in \cite{Cifani:Flandoli25,Luo:Teng25} does not take place.

For comparison, we also investigated the cases when the driving processes, still $\alpha$-stable-like, are either truncated or exponentially tempered. In this case Brownianization takes place. 

We may summarize the intuitions behind the quoted results and the new ones presented here in the following way. Two elements of a turbulence model (of the form said above, namely superposition of divergence-free highly oscillating vector fields weighted by independent stochastic processes) concur to produce a Brownian behaviour of tracers: 
\begin{itemize}
    \item the high number of highly oscillating space components, combined with the independence of the stochastic multipliers. In such a case, the tracer is subject to the effect of several independent inputs, some stronger and some weaker depending on its relative position with respect to the oscillation of the components, changing very rapidly the relative strength of these components, hence restoring some form of chaoticity of the background;
    \item some limitation in the strength or frequency of very large inputs by the stochastic multipliers.
\end{itemize}

In the case of classical $\alpha$-stable cylindrical turbulent fields, the second property above does not hold while the first one is not sufficient to restore a Brownian behaviour. We observe an $\alpha$-stable behaviour of the tracer, namely an anomalous, super-diffusive transport.

\appendix
\section{Scaling parameter for limiting process}
\label{Sect:Appendix}

The main aim of this section is to derive an heuristic formula for the scaling coefficient $\overline{\sigma}$ appearing in Conjecture \ref{main_rmk}. 

We recall that the particle dynamics, when starting from zero, is defined by \eqref{eq:char}:
\[\bX^\eta_t=uC(\eta,\tau,\alpha)\sum_{\bk\in \bK_\eta}\int_0^t\sigma_\bk(\bX^\eta_s) \diamond d{Z}^{\alpha,\bk}_s\]
where, for simplicity, we omitted the dependency from the initial condition $\bO$. We denote by $t_\eta$ the average time the tracer is influenced mostly by a certain Fourier component before changing to another one. It is clear that since for $\eta\ll 1$, more and more Fourier components are added to the dynamics, the average time $t_\eta\ll 1$ is small. Then, it is well-known that one can approximate:
\[\bX^\eta_{t+t_\eta}-\bX^\eta_t \sim uC(\eta,\tau,\alpha)\sum_{\bk\in \bK_\eta}\sigma_\bk(\bX^\eta_t)\left(Z^{\alpha,\bk}_{t+t_\eta}-Z^{\alpha,\bk}_t\right).\]
The validity of such Wong-Zakai approximation of the Marcus stochastic integral has been corroborated by various studies (cf.\ \cite{Marcus78,Kosenkova:Kulik:Pavlyukevich19}).
Since $\bX_t^\eta$ wanders through space and the Fourier modes $\sigma_\bk$ are highly oscillatory, the value of $|\sigma_\bk(\bX_t^\eta)|^\alpha$ is "typically" close to its spatial average:
\[\langle \sigma_\bk\rangle^\alpha := \lim_{R\to+\infty}\int_{\left[-\frac{R}{2};\frac{R}{2}\right]^2}|\sigma_\bk(x)|^\alpha \, d\bx = \frac{1}{2\pi}\int_0^{2\pi} |\sin(t)|^\alpha \, dt \sim \frac{\alpha-1}{\sqrt{\pi}\alpha}\frac{\Gamma(\frac{\alpha-1}{2})}{\Gamma(\frac{\alpha}{2})}.\]
Hence, the tracer mean displacement can be roughly approximated by:
\[\mathbb{E}\left[\left|\bX^\eta_{t+t_\eta}-\bX^\eta_t\right|\right] \sim u\tau^{(\alpha-1)/\alpha}\mathbb{E}\left[\left|Z^{\alpha,\bk}_{t+t_\eta}-Z^{\alpha,\bk}_t\right|\right]\sim u\tau^{(\alpha-1)/\alpha}t_\eta^\frac{1}{\alpha}.\]
Notice now that the increments
\[\bX^\eta_{t_\eta},\bX^\eta_{2t_\eta}-\bX^\eta_{t_\eta},\bX^\eta_{3t_\eta}-\bX^\eta_{2t_\eta},\dots\]
are approximately independent, since each Fourier mode is affected by an independent process. Moreover, each increment has a length $|\bX^\eta_{(i+1)t_\eta}- \bX^\eta_{it_\eta}|$ of order $\lambda\eta$, for a certain parameter $\lambda>0$. Indeed, these increments measure the distance the tracer travel when affected by a certain Fourier component, before "jumping" on another one. Such a typical “distance” to travel in order to jump from one to the other is of the order of the wave-length of the sinusoidal components of the noise, possibly modified by a factor $\lambda$. The intuition behind is that the tracer is on the “top” of a cosine function, where the function takes approximately the value $\pm1$. Moving a little bit, just a portion of the wave-length $\eta$, it will be no more on the top of that cosine component, but more near the top of another component. Summarizing, at each time step $it_\eta$, we can approximate $\bX^\eta_{it_\eta}$ by a random walk with mean displacement of size $i\lambda \eta$.

To conclude our heuristic derivation, we now split the argument between the different cases. When we claim a diffusive limit (i.e.\ cases $(J_2)$ and $(J_3)$) so that approximately
\[\bX^\eta_t\sim \overline{\sigma}\bW_t,\]
for a Brownian motion $\{\bW_t\}_{t\ge0}$ on $\R^2$, the random walk approximation of $\bX^\eta_{it_\eta}$ has finite square-average distance from the origin given by $i (\lambda \eta)^2$. Since it should be also equal to $\overline{\sigma}^2it_\eta$, it follows that
\[\overline{\sigma} \sim \frac{\lambda \eta}{\sqrt{t_\eta}}.\]
By the arguments above, we also have that
\[\lambda\eta \sim u\tau^{(\alpha-1)/\alpha}t_\eta^\frac{1}{\alpha}\]
so that
\[t_\eta \sim \left(\frac{\lambda \eta}{u}\right)^\alpha\tau^{1-\alpha}.\]
Finally, we obtain that
\[\overline{\sigma}\sim  \lambda^{1-\alpha/2}u^{\alpha/2}\eta^{1-\alpha/2}\tau^{(\alpha-1)/2}.\]
Note that by our assumption \eqref{eq:cond}, the parameter $\overline{\sigma}$ is actually independent from $\eta$.

The heuristic derivation for the case $(J_1)$ follows a similar line of reasoning, relying directly on Formula \ref{eq:1} for the mean displacement.

\paragraph{Acknowledgement.}
The research of P.C.\ and L.M.\ is funded by the European Union (ERC, NoisyFluid, No. 101053472). The research of F.F.\ is funded by the European Union (ERC, NoisyFluid, No. 101053472) and the research project PRIN 2022 “Noise in fluid dynamics and related models”, no. 20222YRYSP.

\bibliography{bibli}
\bibliographystyle{alpha}

\end{document}